\documentclass[%
reprint,
superscriptaddress,
amsmath,amssymb,
aps,
prl,
floatfix,
a4paper
]{revtex4-2}

\usepackage{graphicx}% Include figure files
\usepackage{dcolumn}% Align table columns on decimal point
\usepackage{bm}% bold math
\usepackage{hyperref}% add hypertext capabilities
\usepackage{gensymb}
\usepackage{xspace}
\usepackage{textcomp}
\usepackage{xcolor}

\begin{document}

% \preprint{APS/123-QED}

\title{Electronic correlations in promising room-temperature superconductor Pb$_9$Cu(PO$_4$)$_6$O: a DFT+DMFT study}

\author{Dmitry~M.~Korotin}
\email{dmitry@korotin.name}
\affiliation{M.N. Mikheev Institute of Metal Physics of Ural Branch of Russian Academy of Sciences, 18 S. Kovalevskaya St., Yekaterinburg, 620137, Russia.}
\affiliation{Skolkovo Institute of Science and Technology, 3 Nobel St., Moscow, 143026, Russia}

\author{Dmitry~Y.~Novoselov}
\affiliation{M.N. Mikheev Institute of Metal Physics of Ural Branch of Russian Academy of Sciences, 18 S. Kovalevskaya St., Yekaterinburg, 620137, Russia.}
\affiliation{Skolkovo Institute of Science and Technology, 3 Nobel St., Moscow, 143026, Russia}
\affiliation{Department of Theoretical Physics and Applied Mathematics, Ural Federal University, 19 Mira St., Yekaterinburg 620002, Russia}

\author{Alexey~O.~Shorikov}
\affiliation{M.N. Mikheev Institute of Metal Physics of Ural Branch of Russian Academy of Sciences, 18 S. Kovalevskaya St., Yekaterinburg, 620137, Russia.}
\affiliation{Skolkovo Institute of Science and Technology, 3 Nobel St., Moscow, 143026, Russia}
\affiliation{Department of Theoretical Physics and Applied Mathematics, Ural Federal University, 19 Mira St., Yekaterinburg 620002, Russia}

\author{Vladimir~I.~Anisimov}
\affiliation{M.N. Mikheev Institute of Metal Physics of Ural Branch of Russian Academy of Sciences, 18 S. Kovalevskaya St., Yekaterinburg, 620137, Russia.}
\affiliation{Skolkovo Institute of Science and Technology, 3 Nobel St., Moscow, 143026, Russia}
\affiliation{Department of Theoretical Physics and Applied Mathematics, Ural Federal University, 19 Mira St., Yekaterinburg 620002, Russia}

\author{Artem~R.~Oganov}
\affiliation{Skolkovo Institute of Science and Technology, 3 Nobel St., Moscow, 143026, Russia}
\affiliation{Moscow Institute of Physics and Technology, 9 Institutskiy per., Dolgoprudny, Moscow Region, 141701, Russia}

\date{\today}

\begin{abstract}
We present results of the first investigations on the correlated nature of electronic states that cross the Fermi level in Pb$_9$Cu(PO$_4$)$_6$O {\em aka} LK-99 obtained within the DFT + DMFT approach. Coulomb correlations between Cu-$d$ electrons led to the opening of the band gap between the extra-O $p$ and Cu $d_{xz}/d_{yz}$ states. We state that oxygen $p$ states play a significant role in the electronic properties of LK-99. We also assume that doping with electrons is necessary to turn the stoichiometric Pb$_9$Cu(PO$_4$)$_6$O into conducting state.
\end{abstract}

\maketitle

\section{Introduction}
Starting from the first report on the existence of room-temperature superconductivity in Pb$_9$Cu(PO$_4$)$_6$O~\cite{Lee2023} (``LK-99'') there are continuing attempts to clarify which characteristics of the electronic structure could generate the reported compound properties~\cite{Griffin2023,Lai2023,Si2023,oh2023swave}.
It is obviously to suggest, that the $d$-states of the Cu ion in formal $d^9$ electronic configuration are on the Fermi level and the supposed superconductivity is interconnected with them. Following the HTSC cuprates story, one can assume that LK-99 is a system with strong electron-electron correlations, and density functional theory will fail to describe its properties.
DFT+U calculations were already presented in~\cite{Griffin2023, Si2023,Lai2023}.
The insulating band structure was obtained in all three works with the existence of long-range ferromagnetic ordering (the long-range magnetic ordering is the artifact of the DFT+U method). Importance of accounting for correlation effects was reported for many High-T$_c$ superconductors with $d$- or $f$-atoms~\cite{Anisimov2009,Shorikov2020,Weber2010,Lilia2022}. 

In this research we employed the features of DFT+DMFT method to describe electronic structure of strongly correlated paramagnetic systems, taking into account also finite electronic temperature.

\section{\label{sec:methods} Methods}

We performed DFT+DMFT~\cite{DFT+DMFT} calculations following the procedure described in~\cite{hamilt}. On the first step, the DFT calculations  using the Quantum-ESPRESSO~\cite{Giannozzi2009} package with pseudopotentials from the standard solid-state pseudopotential library set~\cite{SSSP} were performed. The exchange-correlation functional was chosen to be in PBEsol form. The energy cut-off for the plane wave function and charge density expansion has been set to 50~Ry and 400~Ry, respectively. Integration in the reciprocal space was done on a regular $4\times4\times5$ $k$-point mesh in the irreducible part of the Brillouin zone. 
The convergence criteria used for crystal cell relaxation within DFT are: total energy $< 10^{-8}$~Ry, total force $< 10^{-4}$~Ry/Bohr, pressure $< 0.2$~kbar.

Next, to take into account Coulomb correlations and many-body effects for the constructed small Hamiltonian, the DFT+DMFT approach~\cite{Anisimov1997,Held2006} was utilized. 
DFT+DMFT calculations were performed at electronic inverse temperature $\beta=1/k_\mathrm{B}T=$40~eV$^{-1}$, where $k_\mathrm{B}$ is the Boltzmann constant and $T$ is the absolute temperature, which corresponds to 290~K. 
An effective DMFT quantum impurity problem~\cite{Werner2006} was solved using the continuous-time quantum Monte Carlo method with the hybridization expansion algorithm~\cite{Gull} as realized in the package \textsc{AMULET}~\cite{AMULET}.

\section{\label{sec:results} Results }
We are still lacking sufficient data regarding the fine crystal structure of the LK-99~\cite{Lee2023} compound. It is known that it was identified as lead apatite Pb$_{10}$(PO$_4$)$_6$O with the P6$_3$/m space group, wherein a copper ion substitutes one of the Pb ions at position $4f$. Consequently, selecting an appropriate crystal structure for electronic structure calculations is not straightforward.

\begin{figure}
	\includegraphics[width=\columnwidth]{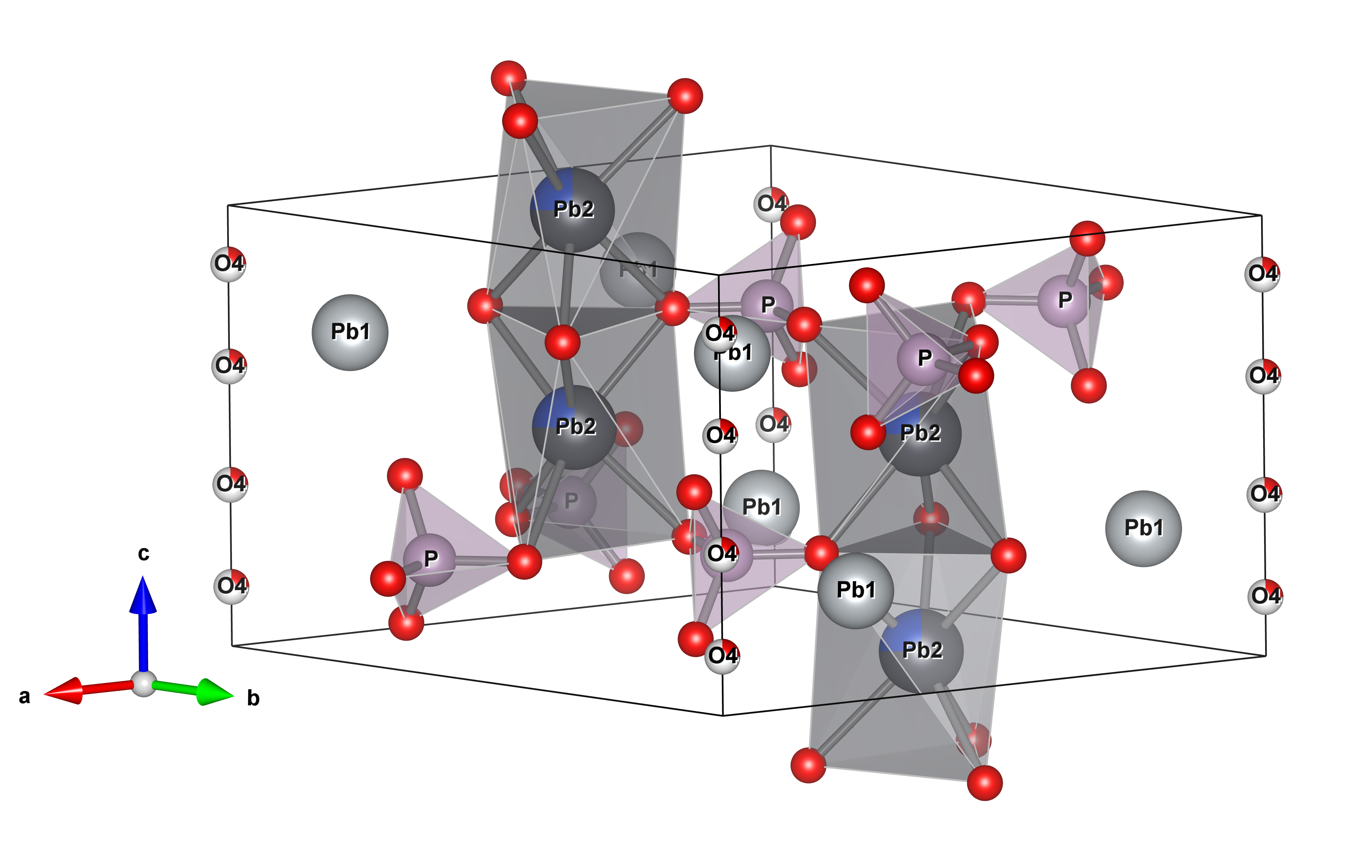}
	\caption{Crystal structure of Pb$_{10}$(PO$_4$)$_6$O. The O4 site (referenced as extra-O in the text) has partial occupation of 0.25. In Pb$_{9}$Cu(PO$_4$)$_6$O copper ion substitutes one of Pb2 ions. Visualized using VESTA~\cite{VESTA}.}
	\label{fig:structure}
\end{figure}

We started our investigation using the Pb$_{10}$(PO$_4$)$_6$O crystal structure~\cite{Krivovichev2003}, as illustrated in Figure~\ref{fig:structure}. It is important to note that the oxygen ion, which is not part of the PO$_4$ octahedra, is located at the Wyckoff site $4e$ with a partial occupation of 0.25. In our analysis, we refer to this oxygen ion as "extra-O" throughout the text.

The problem of dealing with partial occupation sites (or impurity sites) can be approached in two ways: (1) by calculating large supercells with randomly occupied $4e$ sites by oxygen ions, or (2) by staying within a single cell and avoiding relaxation of internal atomic positions of the extra-O (impurity) ion. Avoiding atomic position relaxation in small cells prevents undesirable local distortions that would propagate throughout the crystall due to translation periodicity.

\begin{figure}
	\includegraphics[width=\columnwidth]{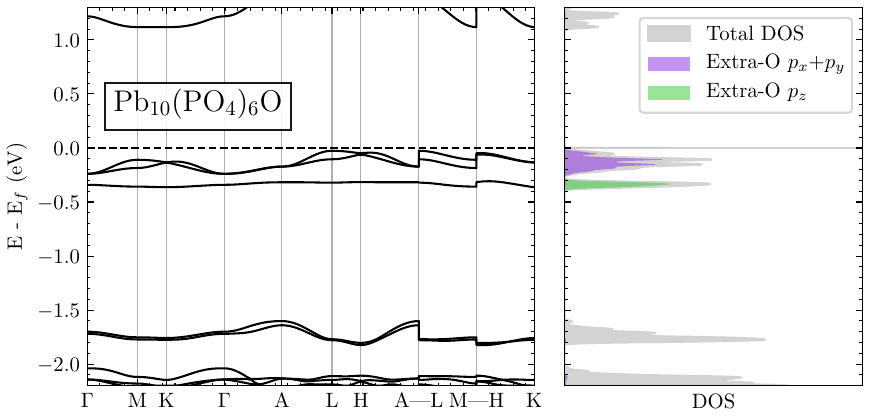}
	\caption{Energy bands (left panel) and densities of states (right panel) for Pb$_{10}$(PO$_4$)$_6$O}
	\label{fig:apatite}
\end{figure}

The full band structure and partial densities of states for the extra-O ion for the parent lead apatite compound are shown in Figure~\ref{fig:apatite}. Calculations were done for the experimental crystal structure with only one extra-O ion in the cell.

The left panel of Figure~\ref{fig:apatite} illustrates the presence of narrow energy bands, with a width of approximately 400~meV, situated just below the Fermi level in Pb$_{10}$(PO$_4$)$_6$O. These three bands arise from the $p$-states of the extra-O ion. The figure provides evidence that the extra-O ion in the lead apatite can be treated as an impurity, exhibiting minimal interaction with the surrounding ions. Consequently, it forms a distinct set of narrow energy bands.

The same rationale was used for a copper ion substitution in LK-99, replacing one of the ten Pb ions. The copper ion acts as an impurity that exerts negligible influence on the average positions of the Pb ions. Consequently, to obtain the crystal cell representing Pb$_{10-x}$Cu$_x$(PO$_4$)$6$O (x=1), we substituted one of the lead ions in Pb$_{10}$(PO$_4$)$_6$O with copper. Subsequently, we fixed the atomic positions of Pb, Cu, and extra-O ions and performed relaxation for all other structural parameters, including the crystal shape, volume, and positions of all remaining atoms.
The resulting lattice parameters $a =  9.748 \AA$, $c =  7.218 \AA, V = 594 \AA^3$ agree with the already published data~\cite{Lee2023, Si2023, Griffin2023}.

\begin{figure}
	\includegraphics[width=\columnwidth]{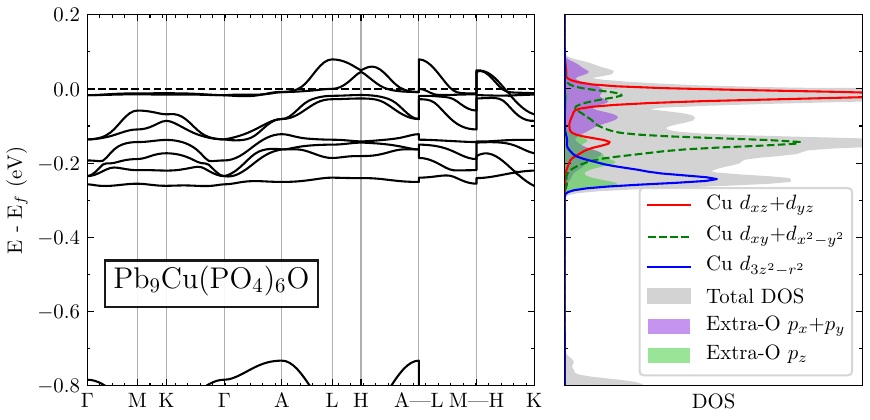}
	\caption{Energy bands (left panel) and densities of states (right panel) for Pb$_9$Cu(PO$_4$)$_6$O}
	\label{fig:pdos}
\end{figure}

The calculated band structure and partial densities of states for the optimized cell of Pb$_9$Cu(PO$_4$)$_6$O are depicted in Figure~\ref{fig:pdos}. 
The copper ion states also form impurity-like energy bands near the Fermi level, in the energy region slightly higher than the extra-O bands. The near-the-Fermi bands set is narrower than in previously presented works~\cite{Griffin2023,Si2023} right because we didn't allow undesirable extra hybridization of Cu and extra-O ions with the nearest-neighbor ions states.

The crystal field of the trigonally distorted oxygen octahedra splits the Cu $d$ -shell into double degenerate subshell corresponding to the irreducible representation $e_g^\sigma$ ($d_{xz}$ and $d_{yz}$ orbitals), double degenerate $e_g^\pi$-subshell ($d_{x^2-y^2}$ and $d_{xy}$ orbitals) and $d_{3z^2-r^2}$ orbital that corresponds to the representation $a_{1g}$. This splitting is clearly seen in partial densities of states in the right panel of Figure~\ref{fig:pdos}.

Two exceptionally narrow energy bands intersecting the Fermi level correspond primarily to electronic states with Cu $d_{xz}$ and $d_{yz}$ orbital symmetries. These partially filled bands have a width of only about 120~meV, suggesting that strong electronic correlations undoubtedly play a crucial role in their behavior. This fact was already accentuated before~\cite{Griffin2023,Si2023}.

The two bands within the energy range of [-0.15, -0.01]~eV below the Fermi level are attributed to the $p_x$ and $p_y$ orbitals of the extra-O ion. Due to the considerable distance between the Cu and extra-O ions (approximately 5.7 $\AA$), there is minimal overlap between the copper ion orbitals and the extra-O $p$-orbitals, resulting in negligible hybridization between them. Below there are also Cu $d_{x^2-y^2}$/$d_{xy}$, extra-O $p_z$ and Cu $d_{3z^2-r^2}$ related energy bands. The width of the entire band set is about 350~meV.

The eight states mentioned above (5 Cu-$d$ and 3 extra-O-$p$) are the minimal basis for the model which could be used for evaluation of the strong Coulomb interaction in Pb$_{10-x}$Cu$_x$(PO$_4$)$6$O. Using the projection procedure (see~\ref{sec:methods} Methods section) we constructed eight Wannier functions with the symmetry of corresponding atomic orbitals. Then the Hamiltonian of the Hubbard model Hamiltonian was constructed in the basis of these Wannier functions. The five Wannier functions with Cu $d$ symmetry were considered corellated states, and the tree O-$p$ like Wannier functions were treated as a bath. 
%The obtained Hamiltonian reproduces the eight closest to the Fermi level energy bands.

\begin{figure}
	\includegraphics[width=\columnwidth]{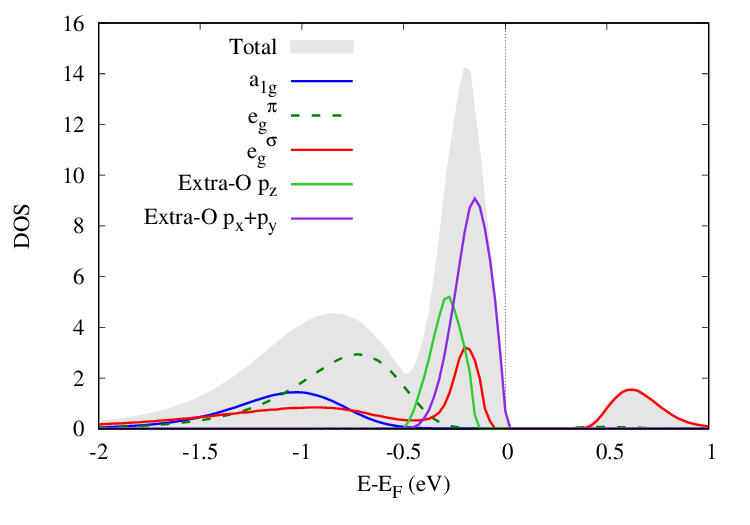}
	\caption{DMFT densities of states for Pb$_9$Cu(PO$_4$)$_6$O.}
	\label{fig:dosDMFT}
\end{figure}

To take into account strong Coulomb correlations in the narrow energy bands we used the approach of Dynamical Mean Field Theory to solve the Hubbard model for the constructed small Hamiltonian. The inverse temperature was $\beta$=40~eV ($\approx$290~K) and Coulomb repulsion parameter $U$=1.8~eV.%, which was the minimum necessary for the appearance of an energy gap in the copper d-shell. 
The obtained spectral functions are shown in Fig.~\ref{fig:dosDMFT}.

Even such a small U value opens the gap about 0.4~eV and results in the insulator solution. 
Electronic states with $e_g^\pi$ and $a_{1g}$ symmetry are completely filled. The spectral function of the $e_g^\sigma$ states demonstrates the appearance of the lower and upper Hubbard bands at -1.1 and 0.6~eV respectively.
The obtained averaged squared magnetic moment is 0.99 $\mu_B^2$ which correspond to 1 hole in doubly degenerate $e_g^{\sigma}$ orbitals. 
The DFT+DMFT results show that taking into account Coulomb correlation is important since it significantly changes the band structure in vicinity of Fermi level obtained in DFT and hence the Fermi surface topology which could be very significant for description superconductivity if it will be confirmed in this compound. 
%(see Figs.~\ref{fig:pdos,fig:dosDMFT}. 
The Figure~\ref{fig:dosDMFT} shows that the top of the valence band is formed by extra-O $p_x + p_y$ states which reminiscent the Pb$_10$(PO$_4$)$_6$O since in case of absent of Cu the top of valence band is also formed by extra-O $p$-states. The spectral functions look similar to charge-transfer insulators, {\em i.e.} NiO or LiNiO$_2$~\cite{Korotin2019}, but the situation is different in Pb$_9$Cu(PO$_4$)$_6$O. In charge transfer insulators the energy gap is formed by metal $d$-states and the nearest ligands $p$-states. In LK-99 the states that cross the Fermi level are $p$-states of extra-O ion which is at least 5.7~\AA away from the copper ion.

We assume that further doping of the system with electrons or holes will lead to it's metallization and the appearance of $e_{g}^{\sigma}$ or $e_{g}^{\pi}$ states at the Fermi level, correspondingly. 
%It can be expected that the introduction of electrons into the original compound leads to the appearance of a room-temperature conductivity of this compound, if it will be confirmed.

\section{Conclusion}
We investigated the electronic structure and the role of correlation effects in Pb$_9$Cu(PO$_4$)$_6$O in frames of DFT+DMFT approach. It was shown that proper accounting for the disorder in occupation of extra-O and Cu/Pb sites in the parent lead apatite structure is especially important and can result in the appearance of narrow energy bands in the vicinity of the Fermi level in the case of the LK-99 compound. These energy bands originate from overlapping of two bands groups, namely Cu-$d$ and extra-O $p$. 
%Both sets of bands could be considered as impurity levels. 
Then using the DFT+DMFT approach, showed that accounting for Coulomb correlations leads to the band gap opening and drastically change band structure obtained in DFT. However, the physical picture
in this compound is much more complicated and cannot be reduced either to Mott insulator on a triangular lattice formed by the Cu $d_{xz}$ and $d_{yz}$ orbitals or charge transfer insulator, due to the remarkable complicated structure of the valence band of Pb$_9$Cu(PO$_4$)$_6$O formed not by the ligands closest to the Cu ion but by distant extra-O $p$ states.

\section*{Acknowledgments}
The DFT parts of the study were supported by the Ministry of Science and Higher Education of the Russian Federation (No.~122021000039-4, theme "Electron").
The DMFT results were obtained within the state assignment of the Russian Science Foundation (Project 19-72-30043).

\bibliographystyle{apsrev4-1}
\bibliography{main}

\end{document}